# Peak effect phenomena, surface superconductivity and paramagnetic Meissner effect in a spherical single crystal of niobium

**Pradip Das[1], C.V. Tomy[1], H. Takeya[2], S. Ramakrishnan[3] and A.K. Grover[3]**

[1]Department of Physics, Indian Institute of Technology Bombay, Mumbai 400076, India.

[2]National Institute of Material Science, Ibaraki 305-0047, Japan.

[3]DCMP&MS, Tata Institute of Fundamental Research, Mumbai 400 005, India.

E-mail: grover@tifr.res.in

**Abstract.** We have explored the vortex phase diagram in a spherical single crystal of niobium ($T_c$ ~ 9.3 K) via isothermal and temperature dependent dc magnetization and ac-susceptibility measurements. The crystal has extremely weak pinning that can be inferred from the reversibility of $M$-$H$ loops below $T_c$. However, one can visualize the peak effect (PE) feature in the isothermal $M$-$H$ loops up to $T$ = 8 K. The PE is also prominent in isothermal ac-susceptibility data for $H$ > 750 Oe. An interesting observation in the present study is the prominent fingerprints of surface superconductivity, starting just above the collapse of pinning at the peak position of the PE and ending at the surface critical field ($H_{c3}$). We have also observed the paramagnetic Meissner effect in field-cooled magnetization data recorded at relatively large fields in this crystal. A vortex phase diagram is constructed by marking the peak positions of the PE ($H_p$), the upper critical field ($H_{c2}$) and the surface critical field ($H_{c3}$). Unlike a previous report which shows the existence of a multi-critical point in the phase diagram of a Nb crystal, where $H_p$, $H_{c2}$ and $H_{c3}$ lines meet, we do not observe a multi-critical point in our weak pinning crystal.

## 1. Introduction

The vortex state of elemental superconductor Nb ($T_c$ ~ 9.3 K) has been receiving attention even in the contemporary times [1-8]. The symmetry of the flux line lattice (FLL) over the entire field-temperature ($H,T$) phase space [1], the range over which flux lines remain correlated [2], order-disorder transition in FLL correlations [2,3,4], surface superconductivity and existence of multi-critical point [4], paramagnetic Meissner effect [5,6,7], etc., continue to get probed extensively in single crystal samples of niobium. In particular, Ling el al. [4] and Park et al. [2] focussed attention on the observation of quintessential peak effect phenomenon in a weakly pinned single crystal of Nb, its intrinsic correlation with the detection of surface superconductivity in the ac susceptibility data and concurrent estimation of azimuthal and radial width of diffraction peaks emanating from the FLL across the PE region. They surmised the existence of a multi-critical point in the ($H,T$) phase diagram where the following lines meet; the order-disorder line *a la* PE phenomenon imbibing supercooling/superheating effects, the $H_{c2}$ (or $T_{c2}$) line following the PE, the $H_{c3}$ (or $T_{c3}$) line located well above $H_{c2}$ line (at low temperatures, $T$ < 8 K) and a mean field like continuous line at lower fields ($H$ < 1000 Oe, $T$ > 8 K); and emphasised the linkage between the termination of the PE line and the $H_{c3}$ line. In view of the notion that the PE reflects pinning due to the quenched random imperfections in the bulk and $H_{c3}$ depends on the surface smoothness and other geometrical parameters of the Nb samples [5,7,9], it is instructive to study the PE phenomenon and the above stated other features in a few other Nb single crystals. We report here on the isofield ac and dc magnetization measurements and the record of magnetization hysteresis in a *spherical* single crystal of niobium having $T_c \sim 9.3$ K. This crystal not only displays the PE phenomenon and surface superconductivity in isofield ac susceptibility plots down to ~750 Oe, but also imprints the PE in isothermal hysteresis ($M$ - $H$) loops and the paramagnetic effects in isofield dc magnetization plots, ranging from 500 Oe to 4000 Oe. The vortex phase diagram in this crystal does not include a multi-critical point, instead the $T_{c2}$ and $T_{c3}$ lines seem to extend up to $T_c(0)$, thereby implying that material characteristics of a given crystal perhaps determine the notion of multi-critical point in elemental Nb.

## 2. Results and discussion

The spherical single crystal (mass = 49.1 mg) of niobium used for the present work was grown in a container-less condition by electrostatic levitation and melting by laser irradiation under high vacuum [10]. Dc magnetization was recorded either on a SQUID-VSM or a PPMS-VSM (Quantum Design, USA) and the ac susceptibility measurements were made using the PPMS-ACMS attachment (Quantum Design, USA). All the magnetic data were recorded for a dc field applied parallel to [100] direction.

Figure 1 shows a portion of the $M$ - $H$ loop recorded at 2 K. The upper inset panel shows an expanded view of the $M$ - $H$ loop encompassing the peak effect (PE) region, just below the closure of the hysteresis loop and approaching the upper critical field, $H_{c2}$. The field value at which the peak of the PE occurs is marked as $H_p$. The value of initial slope corresponding to the virgin (ZFC) portion of the magnetization curve is 1.59, which may be compared with the ratio (1.52) of the maximum magnetization ($-4\pi M$ (max) = 1820 Oe) and the turnover field (~1192 Oe). Both the numbers compare favourably with the ideal 3/2 value expected for a sample with spherical shape.

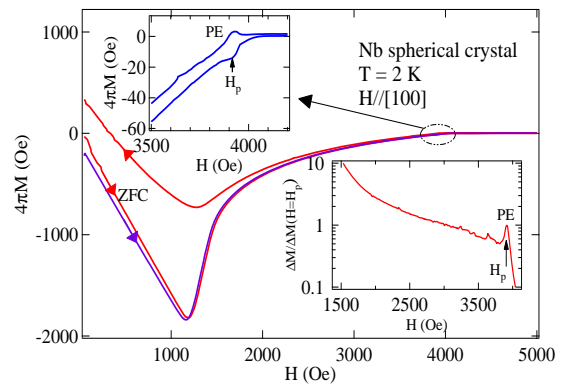

**Figure 1.** Portion of the magnetization hysteresis loop in the 1st, 2nd and 5th quadrants (main panel). The upper inset panel focuses on the field interval of the PE. The lower inset displays a plot of normalized $J_c$ vs. field at 2 K.

Note that the magnetic hysteresis above ~1500 Oe is rather small. Assuming that the residual hysteresis above notional $H_{c1}$ is due to bulk pinning, we show in the lower inset panel of figure 1, the plot of $J_c$ ($\propto \Delta M(H)$) above $H_{c1}$, normalized to the value at $H_p$. The PE indeed is a tiny anomaly; implying a shrinkage of only four times in correlation volume of the FLL from onset to peak of the PE. Note that the bulk pinning collapses as soon as $H$ exceeds $H_p$ and $J_c$ drops to very small values.

Figure 2 shows normalized values of in-phase ($\chi'$) and out of phase ($\chi''$) ac susceptibility ($h_{ac}$ = 2.5 Oe, $f$ = 211 Hz) data at few representative dc fields. In nominal zero dc field, $\chi'$ sharply changes from −1 to 0 across $T_c(0)$ of 9.3 K, whereas $\chi''$ displays a large peak before collapsing to a small value anticipated in the normal state. In large dc fields (e.g., $H$ = 2000 Oe and 1000 Oe, in figure 2), $\chi'$ shows a sharp change across the peak temperature $T_p$ of the PE, followed by a broad slow decline towards the normal value of zero. Following the observation of a similar behaviour by Park et al [2] in a Nb crystal, we identify the $\chi'$ signal, from a little above $T_p$ to the approach to zero, with surface superconductivity. We have marked the values of $T_{c2}$ (approach to upper critical field $H_{c2}$) and $T_{c3}$ (approach to surface critical field $H_{c3}$) in various curves in figure 2. We would like to point out a new feature, a paramagnetic response, followed by a sharp fall in $\chi'$ curve for $H$ = 250 Oe, just before the approach to the respective $T_{c3}$ limit. A similar feature of smaller magnitude can also be noted in the corresponding curve at $H$ = 500 Oe. We believe that the paramagnetic response traces its origin to the existence of paramagnetic Meissner effect (PME) in our sample, as detailed below.

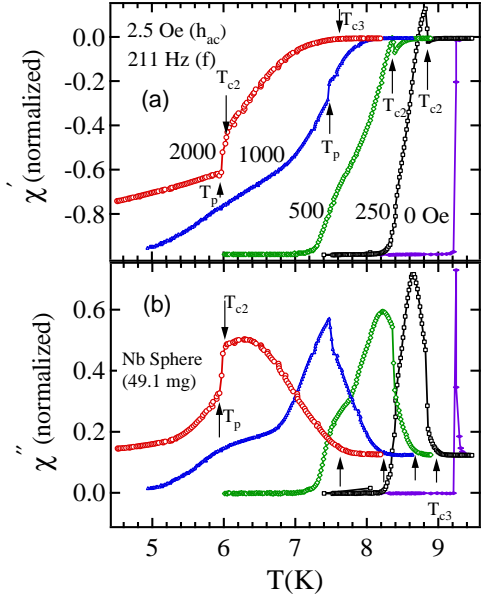

**Figure 2.** Panel (a) and (b) respectively, show the in-phase ($\chi'$) and out of phase ($\chi''$) ac susceptibility data at few selected dc fields. $T_p$, $T_{c2}$, and $T_{c3}$ have been identified in some representative curves.

In order to trace out in detail the signature of PME and surface superconductivity in our crystal, we have measured field-cooled (FC) magnetization while cooling (FCC) as well as heating (FCW) the sample at various applied fields. Figure 3(a) shows portions of FCW magnetization curves for 500 Oe < $H$ < 1500 Oe. The inset panel in figure 3(a) shows the complete $M_{FCW}$ curve for $H$ = 750 Oe. It can be noted that the large saturated diamagnetic response falls towards the zero value as the dc field penetrates the bulk of the sample at $T$ > 6 K, i.e., when the applied field (= 750 Oe) exceeds the temperature-dependent $H_{c1}(T)$ value. However, a closer look at this $M_{FCW}$ curve reveals an anomaly before entering the normal state (see the expanded version of the curve in the main panel of figure

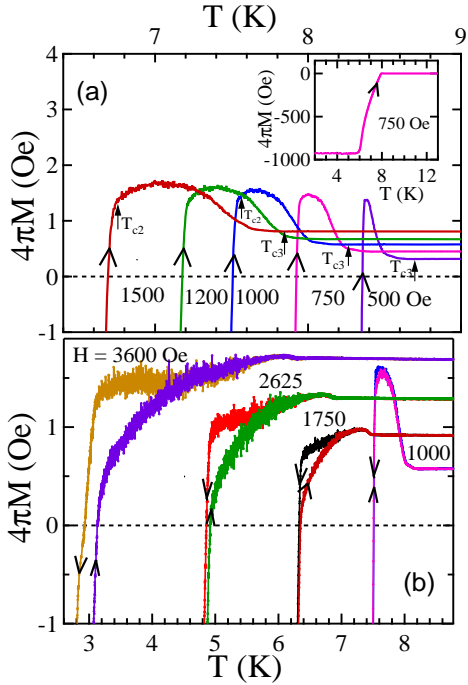

**Figure 3.** Panels (a) and (b) display the temperature dependences of $M_{FCC}(M_{FCW})$ values close to the respective $T_c(H)$ values at different dc fields (as indicated). Arrows indicate the direction in which the data is taken. The inset panel in (a) shows the $M_{FCW}$ curve for $H$ = 750 Oe over the entire temperature range

3(a)). The magnetization value exceeds the corresponding value in the normal state (at $T > T_{c3}$) before reaching the normal state. In view of the often reported [5,7] PME signals in clean samples of Nb, we assign the anomalous increase in magnetization values to the existence of PME in our Nb spherical crystal. It is interesting to find that up to $H$ = 1000 Oe, irreversibility between $M_{FCC}$ and $M_{FCW}$ is negligible. However, for higher fields ($H \geq 1750$ Oe), the irreversibility in the temperature interval of PME gradually increases as $H$ increases (see Fig. 3(b)). It is indeed tempting to seek an intimate connection between the temperature intervals of PME signals in figure 2 and figure 3. Both the intervals at a given field are of the same order. The PME signal is seen to become rather small (hardly discernible) as $H$ exceeds 3600 Oe. Even then, the temperature range it spans can easily be identified (cf. $M_{FCC}$ and $M_{FCW}$ curves for $H$ = 3600 Oe and 2625 Oe in figure 3(b)).

Figure 4 shows a collation of all the $T_p$ (or $H_p$), $T_{c2}$ and $T_{c3}$ values from different magnetization data (ac and dc) in the form of a vortex phase diagram for our spherical crystal of Nb. The inset shows $H_{c1}(T)$ values determined from the dc magnetization data using deviation from linearity criterion for the virgin (ZFC) portion of the M-H curves. In this vortex diagram, we do not find the notion of a multi-critical point where different phase lines could meet, as in the data of Park et al [2]. The observation of PME in our sample, along with its intimate correlation with the surface superconductivity at low fields, indeed implies that the $T_{c3}$ and $T_{c2}$ lines extend up to $T_c$ in nominal zero field. Considering that $H_{c2}(T)$ values in our sample appear smaller than the corresponding values in the sample of ref. [2], we may conclude that the lesser levels of quenched inhomogeneity in our sample along with a characteristic surface morphology perhaps accounts for the observed differences in the vortex phase diagrams of the two specimen of Nb.

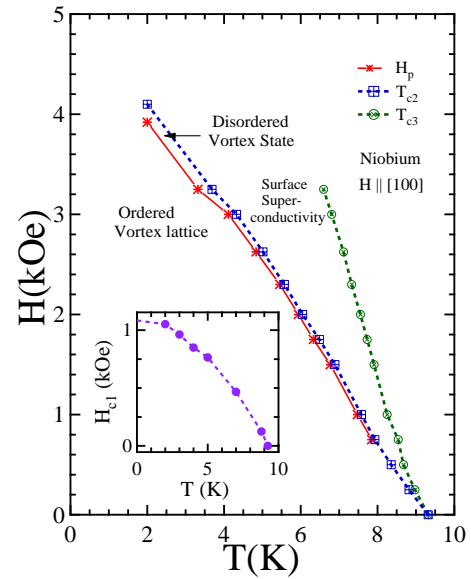

**Figure 4.** Vortex phase diagram for a spherical crystal niobium for $H \parallel [100]$. Different phase boundaries (see text) and vortex phases have been marked. The inset panel displays the plot of $H_{c1}(T)$ for the same sample.

**Acknowledgment**
We would like to very gratefully acknowledge Prof. S.S. Banerjee for many useful discussions about the peak effect phenomenon, surface superconductivity and paramagnetic Meissner effect in superconductors.